\documentclass[a4j,12pt]{article}
\usepackage{amsmath,amsthm,amssymb}
\usepackage{latexsym}
\usepackage{amsmath}
\usepackage{amssymb}
\usepackage{wrapfig}
\usepackage{cases}
\usepackage{fancybox}
\usepackage{bm}
\usepackage{multirow}
\usepackage{graphicx}

\newcommand{\Gcenter}[2]{
	\dimen0=\ht\strutbox%
	\advance\dimen0\dp\strutbox%
	\multiply\dimen0 by#1%
	\divide\dimen0 by2%
	\advance\dimen0 by-.5\normalbaselineskip
	\raisebox{-\dimen0}[0pt][0pt]{#2}}%

\begin{document}
\baselineskip=1.2\baselineskip

\pagestyle{plain}
\begin{center}
{\large \bf Extension of Dynamic Network Biomarker using the propensity score method: Simulation of causal effects on variance and correlation coefficient}

\vspace{1cm}
Satoru Shinoda${}^{1}$, Hideaki Kawaguchi${}^{2}$\textsuperscript{*}\\
\vspace{0.5cm}
${}^{1}${\it Department of Biostatistics, Yokohama City University, School of Medicine, Yokohama City, Kanagawa, 236-0004, Japan}\\
${}^{2}${\it Medical Research Center for Pre-Disease State (Mebyo) AI, Graduate School of Medicine, The University of Tokyo, Bunkyo-ku, Tokyo, 113-0033, Japan}\\
\vspace{0.5cm}
*Corresponding author: Hideaki Kawaguchi\\
E-mail: hikawaguchi@g.ecc.u-tokyo.ac.jp\\
\vspace{0.5cm}
\textbf{Abstract}
\end{center}
In clinical biomarker studies, the Dynamic Network Biomarker (DNB), which is defined based on the variance and Pearson correlation coefficient of biological signals, is often used to detect early signs of disease. 
When applying DNB to clinical data, it is important to account for confounding bias.
However, little attention has been paid to statistical causal inference methods for variance and correlation coefficients.
We first provide a theoretical formulation of causal effects on variance and correlation within the potential outcomes framework, thereby establishing the validity of causal estimation using propensity scores. 
We also evaluate the effectiveness of propensity score matching (PSM) in reducing confounding bias through Monte Carlo simulations. 
Our results highlight the applicability of PSM to DNB analysis and support their use in clinical studies where randomization is not feasible.

\begin{flushleft}
\\
\end{flushleft}

\newpage
\noindent \textbf{\large 1. Introduction}

Recent research has focused on quantitatively characterizing the intermediate state between health and disease, commonly referred to as the ``pre-disease'' state (Chen $et~al.$, 2012; Aihara $et~al.$, 2022). 
The concept of pre-disease dates back to the ancient Chinese medical text ``Huangdi Neijing'' (Chen $et~al.$, 2012; Aihara $et~al.$, 2022) and remains relevant in modern health policy, where health and disease are increasingly viewed as a continuum. 
Since pre-disease can revert to a healthy state with timely intervention but may progress to disease if left undetected, early detection is critical. 
However, due to the lack of a clear definition, rigorous and quantitative investigation of this state remains essential (Aihara $et~al.$, 2022).

Traditional biomarker studies for early disease detection have relied on static features of biological signals, such as mean or instantaneous values. 
However, these approaches have limitations in detecting early signs of deterioration. 
The Dynamic Network Biomarker (DNB) was introduced as a method that focuses on signal variability, such as fluctuations (Chen $et~al.$, 2012). 
DNB has shown useful in identifying early signs of critical transitions in mouse models (Koizumi $et~al.$, 2019), and more recently, its potential has been demonstrated in human data as well (Fang $et~al.$, 2023; Zhang $et~al.$, 2024; Gao $et~al.$, 2024; Zhang $et~al.$, 2023). 
DNB is a simple statistic defined as the product of the standard deviation and the Pearson correlation coefficient of biological signals. 
However, applying DNB to clinical data introduces challenges not present in mouse models, such as individual variability and confounders. 
In particular, adjustment for confounders is essential when comparing DNB values between groups, but previous studies have not sufficiently examined this point.

When using clinical data, type I errors (mistakenly detecting differences when none exist) must be minimized, particularly because DNB is often used in exploratory analyses where numerous biomarker candidates are screened.
The population causal effect on variance, explained as the difference in variances between exposed and unexposed groups, has been introduced in the literature (Hern\'an and Robins, 2020). 
While causal inference for means has been widely studied, little work has focused on causal inference for variance or correlation coefficients. 
For example, a Web of Science search using the terms ``causal effect on the variance,'' ``variance causal inference,'' or ``causal variance'' yielded only 11 results, and none on arXiv. 
Although studies such as Kuroki and Miyakawa (2003) and Tezuka and Kuroki (2023) have proposed point estimation approaches based on Pearl's structural equation models, these are not directly suited for statistical testing in group comparisons. 
We found no studies applying Rubin's framework of propensity scores (PS) to causal inference for variance or Pearson correlation coefficient.

This study focuses on the causal effects of variance and the Pearson correlation coefficient. 
Specifically, we aim to develop a statistical testing framework for group comparisons of DNB components under the potential outcomes framework.
To this end, we first provide a theoretical formulation of the causal effects on variance and correlation, thereby establishing the validity of causal estimation using PSs.
We then examine whether PS matching (PSM) can effectively adjust for confounding in group comparisons. 
Using Monte Carlo simulations, we investigate the degree to which unadjusted analyses may lead to type I error inflation in the absence of true group differences, and explore how the inclusion or exclusion of variables in the PS model affects confounding adjustment.
\\

\newpage
\noindent \textbf{\large 2. Materials and methods}\\
\noindent \textbf{\large 2.1. Causal formulation and PS methodology}\\
\noindent \textbf{\large 2.1.1. Potential outcomes for variance and correlation}

We define the causal effect of variance and that of the Pearson correlation coefficient.
Let $X \in \{0, 1\}$ be a binary exposure variable, and $Y^1$ and $Y^0$ denote the potential outcomes under treatment and control, respectively.
Noting that the observed outcome $Y_i$ is given by $Y_i = X_i Y^1_i + (1 - X_i) Y^0_i$, where $i = 1, \ldots, n$ denotes individuals in the sample.

We define the population causal effect on variance ($\mathrm{PCE_{Var}}$) as the ratio of the variances of the potential outcomes:
\[
\mathrm{PCE_{Var}} = \frac{Var(Y^1)}{Var(Y^0)}.
\]
Similarly, the causal effect on variance among the treated and untreated groups ($\mathrm{PCT_{Var}}$ and $\mathrm{PCU_{Var}}$) are defined as:
\begin{align*}
\mathrm{PCT_{Var}} = \frac{Var(Y^1 \mid X = 1)}{Var(Y^0 \mid X = 1)}, \\
\mathrm{PCU_{Var}} = \frac{Var(Y^1 \mid X = 0)}{Var(Y^0 \mid X = 0)}.
\end{align*}

We also consider the Pearson correlation between two outcome variables, $Y_{[A]}$ and $Y_{[B]}$, which is defined as
\[
Corr(Y_{[A]}, Y_{[B]}) = \frac{Cov(Y_{[A]}, Y_{[B]})}{\sqrt{Var(Y_{[A]}) Var(Y_{[B]})}}.
\]
The causal effects on the Pearson correlation are then defined as:
\begin{align*}
\mathrm{PCE_{Corr}} &= Corr(Y_{[A]}^1, Y_{[B]}^1) - Corr(Y_{[A]}^0, Y_{[B]}^0), \\
\mathrm{PCT_{Corr}} &= Corr(Y_{[A]}^1, Y_{[B]}^1 \mid X = 1) - Corr(Y_{[A]}^0, Y_{[B]}^0 \mid X = 1), \\
\mathrm{PCU_{Corr}} &= Corr(Y_{[A]}^1, Y_{[B]}^1 \mid X = 0) - Corr(Y_{[A]}^0, Y_{[B]}^0 \mid X = 0).
\end{align*}
\\

\noindent \textbf{\large 2.1.2. Assumptions for causal identification}

Our interest lies in inferring causal relationships through comparisons of potential outcomes. 
However, for each individual, potential outcomes cannot be observed, only the outcome under the received treatment is available. 
To infer the unobserved causal effects using observed outcomes, we impose the following assumptions commonly adopted in the potential outcomes framework: positivity, consistency, and exchangeability.

\begin{itemize}
\item \textbf{Positivity} \\
The probability of receiving each treatment level is strictly positive:
\[
\Pr(X = x) > 0.
\]

\item \textbf{Consistency} \\
If an individual receives treatment $X = x$, then the observed outcome equals the potential outcome under that treatment:
\[
X = x \Rightarrow Y_{[j]} = Y_{[j]}^x \quad \text{for } j = A, B.
\]
This assumption implies that the observed outcome corresponds exactly to the potential outcome under the actual treatment received. 
It extends naturally to higher-order moments:
\[
E[(Y_{[j]}^x)^k \mid X = x] = E[(Y_{[j]})^k \mid X = x].
\]
In practice, this assumption is often expressed in terms of expectations:
\[
E[Y_{[j]}^x \mid X = x] = E[Y_{[j]} \mid X = x].
\]

\item \textbf{Exchangeability} \\
This assumption states that the joint potential outcomes are independent of the treatment assignments:
\[
(Y_{[A]}^1, Y_{[A]}^0, Y_{[B]}^1, Y_{[B]}^0) \perp X.
\]
Note that assuming $(Y_{[j]}^1, Y_{[j]}^0) \perp X$ for each outcome separately is not sufficient, because the joint distribution of potential outcomes may still depend on $X$.
As a result, the conditional and marginal distributions of the potential outcomes coincide. For instance, the variances and correlations of potential outcomes satisfy:
\[
Var(Y_{[j]}^x \mid X = 1) = Var(Y_{[j]}^x \mid X = 0) = Var(Y_{[j]}^x),
\]
\[
Corr(Y_{[A]}^x, Y_{[B]}^x \mid X = 1) = Corr(Y_{[A]}^x, Y_{[B]}^x \mid X = 0) = Corr(Y_{[A]}^x, Y_{[B]}^x).
\]
These equalities follow directly from the independence between $X$ and $Y_{[j]}^x$, which allows substitution of the conditional distribution with the marginal distribution:
\[
E[(Y_{[j]}^x)^2 \mid X = x] = \int (y_{[j]}^x)^2 f(y_{[j]}^x \mid x) dy_{[j]}^x = \int (y_{[j]}^x)^2 f(y_{[j]}^x) dy_{[j]}^x = E[(Y_{[j]}^x)^2],
\]
and therefore,
\begin{align*}
Var(Y_{[j]}^x \mid X = x) &= E[ (Y_{[j]}^x)^2 \mid X = x] - (E[Y_{[j]}^x  \mid X = x])^2\\
&= E[ (Y_{[j]}^x)^2 ] - (E[Y_{[j]}^x ])^2\\
&= Var(Y_{[j]}^x).
\end{align*}

Similarly, the correlation coefficients follow:
\begin{align*}
Cov(Y_{[A]}^x, Y_{[B]}^x \mid X) &= E\!\left[ (Y_{[A]}^x - E[Y_{[A]}^x \mid X])(Y_{[B]}^x - E[Y_{[B]}^x \mid X]) \mid X \right] \\
&= E\!\left[ (Y_{[A]}^x - E[Y_{[A]}^x])(Y_{[B]}^x - E[Y_{[B]}^x]) \mid X \right] \\
&= E[Y_{[A]}^x Y_{[B]}^x \mid X] - E[Y_{[A]}^x \mid X] E[Y_{[B]}^x \mid X] \\
&= E[Y_{[A]}^x Y_{[B]}^x] - E[Y_{[A]}^x] E[Y_{[B]}^x] \\
&= Cov(Y_{[A]}^x, Y_{[B]}^x).
\end{align*}
We note that, given the assumption $(Y_{[A]}^1, Y_{[A]}^0, Y_{[B]}^1, Y_{[B]}^0) \perp X$, it follows that $E[Y_{[A]}^x Y_{[B]}^x \mid X] = E[Y_{[A]}^x Y_{[B]}^x]$.
Therefore,
\begin{align*}
Corr(Y_{[A]}^x, Y_{[B]}^x \mid X)
&= \frac{Cov(Y_{[A]}^x, Y_{[B]}^x \mid X)}{\sqrt{Var(Y_{[A]}^x \mid X)\, Var(Y_{[B]}^x \mid X)}} \\
&= \frac{Cov(Y_{[A]}^x, Y_{[B]}^x)}{\sqrt{Var(Y_{[A]}^x)\, Var(Y_{[B]}^x)}} \\
&= Corr(Y_{[A]}^x, Y_{[B]}^x).
\end{align*}
\end{itemize}

Exchangeability holds naturally when the treatment $X$ is randomly assigned, as randomization ensures that $X$ is independent of the potential outcomes. 
However, in non-randomized (observational) settings, we instead consider \textit{conditional exchangeability}, where independence is assumed given a set of observed covariates $Z$:
\[
(Y_{[A]}^1, Y_{[A]}^0, Y_{[B]}^1, Y_{[B]}^0) \perp X \mid Z.
\]
In such cases, we also require a conditional form of the positivity assumption:
\[
\Pr(X = x \mid Z = z) > 0 \quad \text{for all } z \in Z.
\]
This condition is necessary to guarantee sufficient overlap between treatment groups across covariate profiles, and to ensure that every unit has a nonzero probability of receiving each treatment level within strata defined by $Z$.
As a result, the variances and correlations of potential outcomes satisfy:
\[
Var(Y_{[j]}^x \mid X = 1, Z) = Var(Y_{[j]}^x \mid X = 0, Z) = Var(Y_{[j]}^x \mid Z),
\]
\[
Corr(Y_{[A]}^x, Y_{[B]}^x \mid X = 1, Z) = Corr(Y_{[A]}^x, Y_{[B]}^x \mid X = 0, Z) = Corr(Y_{[A]}^x, Y_{[B]}^x \mid Z).
\]

From the above, it is theoretically possible to identify the causal effects on variance and the Pearson correlation coefficient.
\begin{align*}
\mathrm{PCE_{Var}} &= \frac{Var(Y_{[j]}^1)}{Var(Y_{[j]}^0)} = \frac{Var(Y_{[j]} \mid X = 1, Z)}{Var(Y_{[j]} \mid X = 0, Z)},\\
\mathrm{PCE_{Corr}} &= Corr(Y_{[A]}^1, Y_{[B]}^1) - Corr(Y_{[A]}^0, Y_{[B]}^0) \\
&= Corr(Y_{[A]}, Y_{[B]} \mid X = 1, Z) - Corr(Y_{[A]}, Y_{[B]} \mid X = 0, Z).
\end{align*}

This equivalence allows us to estimate $Var(Y^x_{[j]})$ and $Corr(Y^x_{[A]}, Y^x_{[B]})$ using observed outcomes in the treated ($X = 1$) and untreated ($X = 0$) groups, respectively. 
Therefore, population causal effects on variance and correlation can be identified from observed data, validating causal inference on the components of the DNB within the potential outcomes framework.
\\

\noindent \textbf{\large 2.1.3. Justification of PS estimation}

In practice, it is often difficult to achieve conditional exchangeability directly by high-dimensional covariates $Z$.
To address this, Rosenbaum and Rubin (1983) introduced the PS, defined as the conditional probability of receiving the treatment given covariates:
\[
e(Z) = \Pr(X = 1 \mid Z).
\]
A key property of the PS is that it is a balancing score, meaning that conditional on $e(Z)$, the distribution of covariates $Z$ is the same between treated and untreated groups:
\[
X \perp Z \mid e(Z).
\]
As a result, conditional exchangeability given $e(Z)$ holds:
\[
(Y_{[A]}^1, Y_{[A]}^0, Y_{[B]}^1, Y_{[B]}^0) \perp X \mid e(Z).
\]
The positivity assumption is also required:
\[
0 < e(Z) < 1.
\]

Therefore, using the PS, it is theoretically possible to identify the causal effects on variance and the Pearson correlation coefficient.
\begin{align*}
\mathrm{PCE_{Var}} &= \frac{Var(Y_{[j]}^1)}{Var(Y_{[j]}^0)} = \frac{Var(Y_{[j]} \mid X = 1, e(Z))}{Var(Y_{[j]} \mid X = 0, e(Z))},\\
\mathrm{PCE_{Corr}} &= Corr(Y_{[A]}^1, Y_{[B]}^1) - Corr(Y_{[A]}^0, Y_{[B]}^0) \\
&= Corr(Y_{[A]}, Y_{[B]} \mid X = 1, e(Z)) - Corr(Y_{[A]}, Y_{[B]} \mid X = 0, e(Z)).
\end{align*}

In this study, PSM was performed using the \texttt{MatchIt} package in R (R Core Team, 2025; Ho $et~al.$, 2011). 
The \texttt{matchit()} function offers a flexible tool for PS-based matching, allowing treatment and control groups to be balanced by specifying a model formula. 
As a widely used and easily interpretable approach, nearest neighbor matching was employed, with a 1:1 matching ratio and a caliper of 0.2.
Notably, more detailed discussions of PS methods can be found elsewhere (Rosenbaum and Rubin, 1983; D'Agostino, 1998; Joffe and Rosenbaum, 1999).
\\

\noindent \textbf{\large 2.2. Monte Carlo simulation study}

This study considers the extent to which PSM can reduce the type I errors under the presence of confounding, compared to unadjusted. 
Simulation data were generated based on the causal structure (DAG) shown in Figure 1. 
Furthermore, to examine the impact of variable selection for confounders in the PS model, we considered modified DAG assumptions where Z1 remains as a confounder (as in Figure 1), while Z2 is treated as an exposure-only variable, and Z3 as an outcome-only variable.

\begin{figure}[htbp]
\centering
\includegraphics[width=0.8\textwidth]{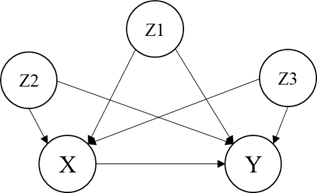}
\caption{Directed acyclic graph for simulation study.}
\end{figure}

We considered three confounders, $Z_1$, $Z_2$, and $Z_3$, each independently sampled from the standard normal distribution $N(0, 1)$.

The exposure variable $X$ given $Z_1$, $Z_2$, and $Z_3$ follows a Bernoulli distribution with a conditional mean given by the function: $\Pr(X = 1 \mid Z_1, Z_2, Z_3) = \Phi(\alpha_1 Z_1 + \alpha_2 Z_2 + \alpha_3 Z_3)$, where $\Phi(\cdot)$ denotes the cumulative distribution function of the standard normal distribution.

The response variables $Y_{[j]}$ (for $j = A, B$) were modeled without a direct effect from $X$, and instead were functions of the confounders $Z_1$, $Z_2$, and $Z_3$, and an error term:
\[
Y_{[j]} = \beta_1 Z_1 + \beta_2 Z_2 + \beta_3 Z_3 + \varepsilon.
\]
The error term $\varepsilon$ follows a normal distribution with mean zero and variance $\sigma^2$, where $\sigma$ depends on $Z_1$, $Z_2$, and $Z_3$, thereby introducing heteroscedasticity:
\[
\sigma = \exp \left( \gamma_1 Z_1 + \gamma_2 Z_2 + \gamma_3 Z_3 \right).
\]
Thus, the model represents a weighted regression framework that does not assume homoscedasticity. 
The two outcomes $Y_{[A]}$ and $Y_{[B]}$ were used to calculate the correlation coefficient within each group.

For each of scenario in the following sections, the sample size was set to $n = 100$, and the number of simulation replications was $S = 100000$.
For variance comparison, we used the \texttt{var.test()} function in R. 
To test differences in Pearson correlation coefficients, we used the \texttt{cocor.indep.groups()} function from the \texttt{cocor} package, which implements Fisher's $z$-transformation method  (Fisher, 1925; Diedenhofen and Musch, 2015), a standard approach in biomarker studies for early disease detection.
\\

\noindent \textbf{\large 2.3. Evaluation of PS model performance}

To evaluate the performance of group comparisons under different levels of confounding adjustment, we analyzed the simulation results using three indicators: (1) Type I error rate, (2) Bias, and (3) Mean squared deviation (MSD).
We also calculated the average post-matching sample size (ASS).

(1) Type I error rate was defined as the proportion of times the null hypothesis that there is no exposure effect was rejected at a two-sided $5\%$ significance level.
This reflects the frequency with which a treatment effect is falsely detected due to apparent variance differences induced by confounding.

(2) Bias was defined as the mean difference between the estimated and true values.
Theoretically, the true value of the variance ratio is 1, and the true difference in Pearson correlation coefficients is 0.
Based on $S$ simulation replications, bias was estimated as follows:
\[
\text{Bias}_v = \frac{1}{S} \sum_{s=1}^S \left( \hat{R}_{v(s)} - 1 \right), \quad
\text{Bias}_c = \frac{1}{S} \sum_{s=1}^S \left( \hat{R}_{c(s)} - 0 \right),
\]
where $\hat{R}_{v(s)}$ denotes the estimated variance ratio and $\hat{R}_{c(s)}$ denotes the estimated difference in Pearson correlation coefficients in the $s$-th simulation replication.

(3) MSD incorporates both the magnitude of bias and the variability of the estimates.
It was calculated as:
\[
\text{MSD}_v = \frac{1}{S} \sum_{s=1}^S \left( \hat{R}_{v(s)} - 1 \right)^2, \quad
\text{MSD}_c = \frac{1}{S} \sum_{s=1}^S \left( \hat{R}_{c(s)} - 0 \right)^2.
\]
\\

\noindent \textbf{\large 2.4. Simulation experiment plan}

To address the study objectives, we designed two sets of simulation scenarios: 
Scenario 1 (Section 2.4.1) evaluates type I error control to exclude markers without causal relationships, while Scenario 2 (Section 2.4.2) explores the impact of variable selection in PS models.
\\

\noindent \textbf{\large 2.4.1. Excluding markers without causal relationships}

This section describes Scenario 1.
Under the null hypothesis of no exposure effect, we considered scenarios based on two frameworks: causal structure (presence or absence of confounding) and error structure (homoscedastic or heteroscedastic). 
The presence or absence of confounding and heteroscedasticity was defined by specific parameter values. 
Confounding was considered absent when $\alpha_1 = \alpha_2 = \alpha_3 = 0$ and $\beta_1 = \beta_2 = \beta_3 = 0$, and present when $\alpha_1 = \alpha_2 = \alpha_3 = 0.25 \text{ or } 0.5$ and $\beta_1 = \beta_2 = \beta_3 = 0.25 \text{ or } 0.5$. 
Similarly, homoscedasticity was defined by setting $\gamma_1 = \gamma_2 = \gamma_3 = 0$, and heteroscedasticity by setting $\gamma_1 = \gamma_2 = \gamma_3 = 0.05 \text{ or } 0.1$.

\begin{itemize}
\item \textbf{Scenario 1.1: No Confounding, Homoscedastic Error} \\
The exposure variable $X$ is generated independently of the confounders $Z_1$, $Z_2$, and $Z_3$, and the outcome variable $Y_{[j]}$ is unaffected by $Z_1$, $Z_2$, and $Z_3$ ($j = A, B$). 
In addition, the error variance $\sigma^2$ is constant. 
This scenario represents an idealized setting that mimics a completely randomized design.

\item \textbf{Scenario 1.2: Confounding Present, Homoscedastic Error} \\
In this setting, $X$ is influenced by $Z_1$, $Z_2$, and $Z_3$, and $Y_{[j]}$ is also affected by $Z_1$, $Z_2$, and $Z_3$, leading to confounding. 
However, the error variance remains constant, indicating that variability in the observed outcomes is not dependent on $Z_1$, $Z_2$, and $Z_3$.

\item \textbf{Scenario 1.3: Confounding Present, Heteroscedastic Error} \\
Both $X$ and $Y_{[j]}$ are influenced by the confounders $Z_1$, $Z_2$, and $Z_3$, and in addition, the error variance varies depending on $Z_1$, $Z_2$, and $Z_3$, introducing heteroscedasticity. 
This scenario reflects realistic variability observed in biomarkers and other clinical measures.

\item \textbf{Scenario 1.4: No Confounding, Heteroscedastic Error} \\
While there is no confounding between $X$ and $Y_{[j]}$, the error variance of $Y_{[j]}$ depends on $Z_1$, $Z_2$, and $Z_3$, creating a heteroscedastic structure. 
Although a rare case in clinical studies, this scenario is included to explore the possibility that confounders affect variance estimation without influencing the mean.
\end{itemize}
Table 1 summarizes the combinations of each scenario and parameter values.
\begin{table}[h]
\centering
\caption{Simulation Settings for Scenario 1}
\begin{tabular}{lccc}\hline
\textbf{Scenario} & $\alpha_i$ & $\beta_i$ & $\gamma_i$ \\\hline
1.1    & 0.00 & 0.00 & 0.00 \\\hline
1.2    &  &  &  \\
1.2.1  & 0.25 & 0.25 & 0.00 \\
1.2.2  & 0.25 & 0.50 & 0.00 \\
1.2.3  & 0.50 & 0.25 & 0.00 \\
1.2.4  & 0.50 & 0.50 & 0.00 \\\hline
1.3    &  &  &  \\
1.3.1  & 0.25 & 0.25 & 0.05 \\
1.3.2  & 0.25 & 0.25 & 0.10 \\
1.3.3  & 0.25 & 0.50 & 0.05 \\
1.3.4  & 0.25 & 0.50 & 0.10 \\
1.3.5  & 0.50 & 0.25 & 0.05 \\
1.3.6  & 0.50 & 0.25 & 0.10 \\
1.3.7  & 0.50 & 0.50 & 0.05 \\
1.3.8  & 0.50 & 0.50 & 0.10 \\\hline
1.4    &  &  &  \\
1.4.1  & 0.00 & 0.00 & 0.05 \\
1.4.2  & 0.00 & 0.00 & 0.10 \\\hline
\end{tabular}

\vspace{1mm}
{\footnotesize \textit{Note}: $\alpha_i$: path from $Z_i$ to $X$, $\beta_i$: path from $Z_i$ to $Y_{[j]}$, $\gamma_i$: parameter influencing the variance of $Y_{[j]}$ from $Z_i$ ($i=1,2,3; j=A,B$).}
\end{table}

\newpage
To examine the impact of confounding adjustment, we compared the following four models:
\begin{enumerate}
\item \textbf{Unadjusted}: No consideration of confounders in group comparisons
\item \textbf{PSM1}: Matching based on a PS estimated using only $Z_1$
\item \textbf{PSM2}: PS estimated using $Z_1$ and $Z_2$
\item \textbf{PSM3}: PS estimated using all three confounders (ideal adjustment PS model)
\end{enumerate}

\newpage
\noindent \textbf{\large 2.4.2. Variable selection for PS models}

This section describes Scenario 2, which considers the impact of variable selection in PS models by changing the causal structures from Scenario 1.
In Scenario 1.2 and 1.3, all of $Z_1$, $Z_2$, and $Z_3$ were assumed to be confounders. However, to explore variable selection perspectives, we extended the scenarios by modifying the assumed structure of the DAG. 
Specifically, we considered alternative structures in which variables functioned solely as confounders, as predictors of the exposure, or as predictors of the outcome. 
The parameter values for each scenario are summarized in Table 2.

\begin{table}[h]
\centering
\caption{Simulation Settings for Scenario 2}
\begin{tabular}{lcccc}\hline
\textbf{Scenario} & $\alpha_1$, $\alpha_2$ & $\beta_1$, $\beta_3$ & $\gamma_1$, $\gamma_3$ & $\alpha_3$, $\beta_2$, $\gamma_2$ \\\hline
2.1 & 0.25 & 0.25 & 0.10 & 0.00 \\
2.2 & 0.25 & 0.50 & 0.10 & 0.00 \\
2.3 & 0.50 & 0.25 & 0.10 & 0.00 \\
2.4 & 0.50 & 0.50 & 0.10 & 0.00 \\\hline
\end{tabular}
\end{table}

To examine the variable selection, we compared the following PS models:
\begin{enumerate}
\item \textbf{PSM1.1}: PS estimated using only $Z_1$
\item \textbf{PSM1.2}: PS estimated using only $Z_2$
\item \textbf{PSM1.3}: PS estimated using only $Z_3$
\item \textbf{PSM2.1}: PS estimated using $Z_1$ and $Z_2$
\item \textbf{PSM2.2}: PS estimated using $Z_1$ and $Z_3$
\item \textbf{PSM2.3}: PS estimated using $Z_2$ and $Z_3$
\item \textbf{PSM3}: PS estimated using $Z_1$, $Z_2$ and $Z_3$
\end{enumerate}

\newpage
\noindent \textbf{\large 3. Results}\\
\noindent \textbf{\large 3.1. Excluding markers without causal relationships}

Tables 3 and 4 summarize the simulation results under Scenario 1 in Section 2.4.1: Table 3 shows variances, and Table 4 shows Pearson correlation coefficients.

Table 3 summarizes the results for variance under Scenario 1.
In the absence of confounding (Scenario 1.1), all approaches maintained type I error rates close to the nominal 5\%, indicating the validity of the simulation design. 
When confounding was introduced (Scenarios 1.2–1.3), unadjusted analyses exhibited inflated type I errors and increasing bias, particularly under heteroscedastic error structures. 
In contrast, PSM substantially reduced both bias and MSD, with a monotonic improvement from PSM1 to PSM3. 
These results confirm that increasing the number of true confounders included in the PS model leads to more accurate estimation and proper control of false positives. 
Although some residual bias remained under strong confounding and heteroscedasticity, the degree of error was markedly lower after PSM adjustment.

Table 4 presents the results for the Pearson correlation coefficient under Scenario 1.
The results exhibited the same overall trend as those for variance.
Compared with the variance results, the magnitude of bias and type I error inflation was smaller, indicating that correlation-based measures were slightly more stable.

\begin{table}[htbp]
\centering
\caption{Results for variance in Scenario 1 (Part 1)}\label{tab:sim1}
\begin{tabular}{lccccc}\hline
Scenario 	& Approach 		& ASS 	& Type I error rate	& Bias 	& MSD \\ \hline
1.1		& Unadjusted	& -- 		& 0.0488 		& 0.0425 	& 0.0972 \\
		& PSM1		& 77.6787	& 0.0492 		& 0.0557 	& 0.1328 \\
		& PSM2		& 76.9290	& 0.0499 		& 0.0559 	& 0.1344 \\
		& PSM3		& 76.1053	& 0.0493 		& 0.0561 	& 0.1359 \\ \hline
1.2.1		& Unadjusted	& --		& 0.0505 		& 0.0434	& 0.0983 \\
		& PSM1		& 75.0170	& 0.0513 		& 0.0598 	& 0.1406 \\
		& PSM2		& 71.2588	& 0.0501 		& 0.0665 	& 0.1501 \\
		& PSM3		& 67.3665	& 0.0489 		& 0.0740 	& 0.1631 \\
1.2.2		& Unadjusted	& --		& 0.0510 		& 0.0438 	& 0.0984 \\
		& PSM1		& 75.0187	& 0.0492 		& 0.0621 	& 0.1398 \\
		& PSM2		& 71.2587	& 0.0458 		& 0.0730 	& 0.1485 \\
		& PSM3		& 67.3673	& 0.0424 		& 0.0878	& 0.1582 \\
1.2.3		& Unadjusted	& --		& 0.0515 		& 0.0432 	& 0.0987 \\
		& PSM1		& 69.9918	& 0.0512 		& 0.0655 	& 0.1539 \\
		& PSM2		& 60.7733	& 0.0504 		& 0.0828 	& 0.1906 \\
		& PSM3		& 51.4233	& 0.0504 		& 0.1101 	& 0.2521 \\
1.2.4		& Unadjusted	& --		& 0.0508 		& 0.0440 	& 0.0978 \\
		& PSM1		& 69.9800	& 0.0500 		& 0.0713 	& 0.1555 \\
		& PSM2		& 60.7432	& 0.0491 		& 0.0985 	& 0.1961 \\
		& PSM3		& 51.3591	& 0.0479 		& 0.1536 	& 0.2746 \\ \hline
1.3.1		& Unadjusted	& --		& 0.0674 		& 0.1477 	& 0.1429 \\
		& PSM1		& 75.0170	& 0.0600 		& 0.1388 	& 0.1827 \\
		& PSM2		& 71.2588	& 0.0545 		& 0.1136 	& 0.1762 \\
		& PSM3		& 67.3665	& 0.0510 		& 0.0864 	& 0.1724 \\
1.3.2		& Unadjusted	& --		& 0.1186 		& 0.2686 	& 0.2368 \\
		& PSM1		& 75.0187	& 0.0847 		& 0.2293 	& 0.2601 \\
		& PSM2		& 71.2587	& 0.0691 		& 0.1683	& 0.2226 \\
		& PSM3		& 67.3673	& 0.0585 		& 0.1402	& 0.1917 \\
1.3.3		& Unadjusted	& --		& 0.0595 		& 0.1154 	& 0.1255 \\
		& PSM1		& 75.0058	& 0.0544 		& 0.1189 	& 0.1674 \\
		& PSM2		& 71.2824	& 0.0494 		& 0.1062 	& 0.1652 \\
		& PSM3		& 67.4071	& 0.0446 		& 0.0944 	& 0.1643 \\
1.3.4		& Unadjusted	& --		& 0.0905 		& 0.1998 	& 0.1832 \\
		& PSM1		& 74.9819	& 0.0724 		& 0.1842 	& 0.2208 \\
		& PSM2		& 71.2648	& 0.0626 		& 0.1470 	& 0.2012 \\
		& PSM3		& 67.3569	& 0.0520 		& 0.1116 	& 0.1844 \\ \hline
\end{tabular}
\end{table}
\addtocounter{table}{-1}
\begin{table}[htbp]
\centering
\caption{Results for variance in Scenario 1 (Part 2)}
\begin{tabular}{lccccc}\hline
Scenario 	& Approach 		& ASS 	& Type I error rate	& Bias 	& MSD \\ \hline
1.3.5		& Unadjusted	& --		& 0.0901 		& 0.2269 	& 0.1889 \\
		& PSM1		& 70.0006	& 0.0687 		& 0.2088 	& 0.2413 \\
		& PSM2		& 60.7432	& 0.0580 		& 0.1707 	& 0.2486 \\
		& PSM3		& 51.4138	& 0.0515 		& 0.1288 	& 0.2690 \\
1.3.6		& Unadjusted	& --		& 0.2090 		& 0.4427 	& 0.4006 \\
		& PSM1		& 69.9871	& 0.1217 		& 0.3702 	& 0.4090 \\
		& PSM2		& 60.7648	& 0.0795 		& 0.2650 	& 0.3426 \\
		& PSM3		& 51.4310	& 0.0568 		& 0.1479 	& 0.2930 \\
1.3.7		& Unadjusted	& --		& 0.0712 		& 0.1735 	& 0.1545 \\
		& PSM1		& 69.9993	& 0.0609 		& 0.1723 	& 0.2138 \\
		& PSM2		& 60.7658	& 0.0549 		& 0.1606 	& 0.2372 \\
		& PSM3		& 51.3882	& 0.0501 		& 0.1645 	& 0.2882 \\
1.3.8		& Unadjusted	& --		& 0.1408 		& 0.3282 	& 0.2797 \\
		& PSM1		& 70.0311	& 0.0951 		& 0.2949 	& 0.3275 \\
		& PSM2		& 60.7488	& 0.0710 		& 0.2369 	& 0.3105 \\
		& PSM3		& 51.4392	& 0.0559 		& 0.1839 	& 0.3210 \\ \hline
1.4.1		& Unadjusted	& --		& 0.0534 		& 0.0443 	& 0.1020 \\
		& PSM1		& 77.6787	& 0.0539 		& 0.0579 	& 0.1382 \\
		& PSM2		& 76.9290	& 0.0535 		& 0.0575 	& 0.1393 \\
		& PSM3		& 76.1053	& 0.0524 		& 0.0577 	& 0.1403 \\
1.4.2		& Unadjusted	& --		& 0.0707 		& 0.0519 	& 0.1182 \\
		& PSM1		& 77.6362	& 0.0669 		& 0.0658 	& 0.1566 \\
		& PSM2		& 76.9611	& 0.0661 		& 0.0662 	& 0.1567 \\
		& PSM3		& 76.1330	& 0.0646 		& 0.0674 	& 0.1578 \\ \hline
\end{tabular}
\end{table}

\begin{table}[htbp]
\centering
\caption{Results for Pearson correlation coefficient in Scenario 1 (Part 1)}\label{tab:sim1}
\begin{tabular}{lccccc}\hline
Scenario 	& Approach 		& ASS 	& Type I error rate	& Bias 	& MSD \\ \hline
1.1 		& Unadjusted	& --		& 0.0495 		& 0.0009 	& 0.0411 \\
		& PSM1		& 77.6787	& 0.0492 		& 0.0008 	& 0.0528 \\
		& PSM2		& 76.9290	& 0.0507 		& 0.0006 	& 0.0538 \\
		& PSM3		& 76.1053	& 0.0496 		& 0.0010 	& 0.0541 \\ \hline
1.2.1		& Unadjusted	& --		& 0.0498 		& -0.0008 	& 0.0396 \\
		& PSM1		& 75.0170	& 0.0495 		& 0.0008 	& 0.0531 \\
		& PSM2		& 71.2588	& 0.0484 		& 0.0022 	& 0.0561 \\
		& PSM3		& 67.3665	& 0.0488 		& 0.0056 	& 0.0598 \\
1.2.2		& Unadjusted	& --		& 0.0495 		& 0.0008 	& 0.0295 \\
		& PSM1		& 75.0187	& 0.0489 		& 0.0027 	& 0.0405 \\
		& PSM2		& 71.2587	& 0.0446 		& 0.0083 	& 0.0425 \\
		& PSM3		& 67.3673	& 0.0409 		& 0.0167 	& 0.0446 \\
1.2.3		& Unadjusted	& --		& 0.0486 		& 0.0004 	& 0.0398 \\
		& PSM1		& 69.9918	& 0.0494 		& 0.0022 	& 0.0575 \\
		& PSM2		& 60.7733	& 0.0487 		& 0.0064 	& 0.0668 \\
		& PSM3		& 51.4233	& 0.0493 		& 0.0167 	& 0.0809 \\
1.2.4		& Unadjusted	& --		& 0.0502 		& 0.0003 	& 0.0322 \\
		& PSM1		& 69.9800	& 0.0498 		& 0.0043 	& 0.0474 \\
		& PSM2		& 60.7432	& 0.0476 		& 0.0147 	& 0.0565 \\
		& PSM3		& 51.3591	& 0.0459 		& 0.0417 	& 0.0697 \\ \hline
1.3.1		& Unadjusted	& --		& 0.0529 		& -0.0143 	& 0.0407 \\
		& PSM1		& 75.0170	& 0.0527 		& -0.0086 	& 0.0543 \\
		& PSM2		& 71.2588	& 0.0513 		& -0.0033 	& 0.0572 \\
		& PSM3		& 67.3665	& 0.0515 		& 0.0040 	& 0.0609 \\
1.3.2		& Unadjusted	& --		& 0.0631 		& -0.0246 	& 0.0439 \\
		& PSM1		& 75.0187	& 0.0605 		& -0.0165 	& 0.0581 \\
		& PSM2		& 71.2587	& 0.0571 		& -0.0053  	& 0.0603 \\
		& PSM3		& 67.3673	& 0.0568 		& 0.0055  	& 0.0637 \\
1.3.3		& Unadjusted	& --		& 0.0542 		& -0.0260 	& 0.0308 \\
		& PSM1		& 75.0058	& 0.0507 		& -0.0162 	& 0.0416 \\
		& PSM2		& 71.2824	& 0.0478 		& -0.0039 	& 0.0435 \\
		& PSM3		& 67.4071	& 0.0431 		& 0.0126 	& 0.0456 \\
1.3.4		& Unadjusted	& --		& 0.0697 		& -0.0158 	& 0.0350 \\
		& PSM1		& 74.9819	& 0.0609 		& -0.0356 	& 0.0452 \\
		& PSM2		& 71.2648	& 0.0551 		& -0.0150 	& 0.0467 \\
		& PSM3		& 67.3569	& 0.0478 		& 0.0101 	& 0.0484 \\ \hline
\end{tabular}
\end{table}
\addtocounter{table}{-1}
\begin{table}[htbp]
\centering
\caption{Results for Pearson correlation coefficient in Scenario 1 (Part 2)}
\begin{tabular}{lccccc}\hline
Scenario 	& Approach 		& ASS 	& Type I error rate	& Bias 	& MSD \\ \hline
1.3.5		& Unadjusted	& --		& 0.0528 		& -0.0198 	& 0.0413 \\
		& PSM1		& 70.0006	& 0.0519 		& -0.0122 	& 0.0592 \\
		& PSM2		& 60.7432	& 0.0516 		& -0.0029 	& 0.0688 \\
		& PSM3		& 51.4138	& 0.0505 		& 0.0127 	& 0.0823 \\
1.3.6		& Unadjusted	& --		& 0.0640 		& -0.0375 	& 0.0448 \\
		& PSM1		& 69.9871	& 0.0593 		& -0.0252 	& 0.0628 \\
		& PSM2		& 60.7648	& 0.0580 		& -0.0101 	& 0.0724 \\
		& PSM3		& 51.4310	& 0.0568 		& 0.0116 	& 0.0861 \\
1.3.7		& Unadjusted	& --		& 0.0588 		& -0.0413 	& 0.0345 \\
		& PSM1		& 69.9993	& 0.0520 		& -0.0268 	& 0.0487 \\
		& PSM2		& 60.7658	& 0.0500 		& -0.0039 	& 0.0574 \\
		& PSM3		& 51.3882	& 0.0468 		& 0.0372  	& 0.0708 \\
1.3.8		& Unadjusted	& --		& 0.0824 		& -0.0813	& 0.0411 \\
		& PSM1		& 70.0311	& 0.0654 		& -0.0560	& 0.0540 \\
		& PSM2		& 60.7488	& 0.0543 		& -0.0204 	& 0.0600 \\
		& PSM3		& 51.4392	& 0.0520 		& 0.0337  	& 0.0736 \\ \hline
1.4.1		& Unadjusted	& --		& 0.0523 		& 0.0090 	& 0.0421 \\
		& PSM1		& 77.6787	& 0.0520 		& 0.0009 	& 0.0541 \\
		& PSM2		& 76.9290	& 0.0537 		& 0.0007 	& 0.0550 \\
		& PSM3		& 76.1053	& 0.0527 		& 0.0011 	& 0.0553 \\
1.4.2		& Unadjusted	& --		& 0.0639 		& -0.0004 	& 0.0461 \\
		& PSM1		& 77.6362	& 0.0630 		& -0.0002 	& 0.0590 \\
		& PSM2		& 76.9611	& 0.0640 		& -0.0001 	& 0.0596 \\
		& PSM3		& 76.1330	& 0.0635 		& -0.0002 	& 0.0601 \\ \hline
\end{tabular}
\end{table}

\newpage
\noindent \textbf{\large 3.2. Variable selection for PS models}

Tables 5 and 6 summarize the simulation results under Scenario 2 in Section 2.4.2: Table 5 shows variances, and Table 6 shows Pearson correlation coefficients.
As the results for variance and Pearson correlation coefficient exhibited similar trends, the following results primarily focuses on the variance results presented in Table 5.

When the true confounder $Z_1$ was included in the PS model (e.g., PSM1.1, PSM2.1, PSM2.2, and PSM3), both type I error and bias remained close to nominal levels for variance and correlation comparisons, confirming adequate control of confounding. 

The type I error control suggested a difference between the roles of $Z_2$ (the exposure-related variable) and $Z_3$ (the outcome-related variable).
Models including only the exposure-related variable $Z_2$ (e.g., PSM1.2 and PSM2.1) showed higher type I error rates, whereas those including the outcome-related variable $Z_3$ (e.g., PSM1.3 and PSM2.2) tended to yield relatively better control of type I error.
These patterns imply that adding exposure-only variables such as $Z_2$ may not improve, and could even distort, the estimation of the PS, while incorporating outcome-related variables like $Z_3$ offers improvement.

Furthermore, the comparison between PSM2.2 and PSM3 suggests that simply including more variables in the PS model does not necessarily lead to better adjustment.
Therefore, careful consideration of causal relationships, for example through causal diagrams such as DAGs, is essential when specifying PS models.

\begin{table}[htbp]
\centering
\caption{Results for variance in Simulation 2}
\begin{tabular}{lccccc}\hline
Scenario 	& Approach		& ASS 	& Type I error rate	& Bias 	& MSD \\ \hline
2.1		& PSM1.1		& 74.8410	& 0.0596 		& 0.0780 	& 0.1588 \\
		& PSM1.2		& 74.8674	& 0.0671 		& 0.1423 	& 0.1970 \\
		& PSM1.3		& 77.6459	& 0.0643 		& 0.1372 	& 0.1818 \\
		& PSM2.1		& 70.9301	& 0.0607 		& 0.0798 	& 0.1712 \\
		& PSM2.2		& 74.0109	& 0.0582 		& 0.0770 	& 0.1585 \\
		& PSM2.3		& 74.0162	& 0.0647 		& 0.1433 	& 0.1958 \\
		& PSM3		& 70.0509	& 0.0590 		& 0.0795 	& 0.1711 \\ \hline
2.2		& PSM1.1		& 74.8561	& 0.0560 		& 0.0762 	& 0.1546 \\
		& PSM1.2		& 74.8773	& 0.0637 		& 0.1208 	& 0.1824 \\
		& PSM1.3		& 77.6458	& 0.0600 		& 0.1192 	& 0.1676 \\
		& PSM2.1		& 70.9510	& 0.0580 		& 0.0771 	& 0.1681 \\
		& PSM2.2		& 74.0178	& 0.0549 		& 0.0758 	& 0.1551 \\
		& PSM2.3		& 74.0552	& 0.0615 		& 0.1222 	& 0.1827 \\
		& PSM3		& 70.0497	& 0.0566 		& 0.0780 	& 0.1688 \\ \hline
2.3		& PSM1.1		& 68.6030	& 0.0597 		& 0.0875 	& 0.1810 \\
		& PSM1.2		& 68.6286	& 0.0761 		& 0.2173 	& 0.2669 \\
		& PSM1.3		& 77.6452	& 0.0734 		& 0.1933 	& 0.2154 \\
		& PSM2.1		& 57.9343	& 0.0592 		& 0.0993 	& 0.2279 \\
		& PSM2.2		& 67.7502	& 0.0577 		& 0.0874 	& 0.1817 \\
		& PSM2.3		& 67.7401	& 0.0748 		& 0.2193 	& 0.2625 \\
		& PSM3		& 57.0990	& 0.0589 		& 0.1001 	& 0.2311 \\ \hline
2.4		& PSM1.1		& 68.5921	& 0.0569 		& 0.0902 	& 0.1771 \\
		& PSM1.2		& 68.5909	& 0.0694 		& 0.1847 	& 0.2403 \\
		& PSM1.3		& 77.6270	& 0.0660 		& 0.1656 	& 0.1938 \\
		& PSM2.1		& 57.8562	& 0.0580 		& 0.1044 	& 0.2299 \\
		& PSM2.2		& 67.7485	& 0.0563 		& 0.0912 	& 0.1798 \\
		& PSM2.3		& 67.7273	& 0.0687 		& 0.1848 	& 0.2420 \\
		& PSM3		& 57.0425	& 0.0572 		& 0.1042 	& 0.2325 \\ \hline
\end{tabular}
\end{table}

\begin{table}[htbp]
\centering
\caption{Results for Pearson correlation coefficient in Simulation 2}
\begin{tabular}{lccccc}\hline
Scenario 	& Approach		& ASS 	& Type I error rate	& Bias	& MSD \\ \hline
2.1		& PSM1.1		& 74.8410	& 0.0569 		& -0.0006 	& 0.0576 \\
		& PSM1.2		& 74.8674	& 0.0584 		& -0.0078 	& 0.0576 \\
		& PSM1.3		& 77.6459	& 0.0560 		& -0.0067 	& 0.0551 \\
		& PSM2.1		& 70.9301	& 0.0576 		& -0.0001 	& 0.0612 \\
		& PSM2.2		& 74.0109	& 0.0571 		& 0.0002  	& 0.0582 \\
		& PSM2.3		& 74.0162	& 0.0578 		& -0.0075	& 0.0583 \\
		& PSM3		& 70.0509	& 0.0577 		& -0.0004	& 0.0619 \\ \hline
2.2		& PSM1.1		& 74.8561	& 0.0536 		& 0.0001 	& 0.0479 \\
		& PSM1.2		& 74.8773	& 0.0576 		& -0.0171	& 0.0480 \\
		& PSM1.3		& 77.6458	& 0.0547 		& -0.0141 	& 0.0463 \\
		& PSM2.1		& 70.9301	& 0.0562 		& 0.0010 	& 0.0510 \\
		& PSM2.2		& 74.0178	& 0.0529 		& 0.0014 	& 0.0484 \\
		& PSM2.3		& 74.0552	& 0.0565 		& -0.0158 	& 0.0484 \\
		& PSM3		& 70.0497	& 0.0544 		& 0.0011 	& 0.0514 \\ \hline
2.3		& PSM1.1		& 68.6030	& 0.0566 		& 0.0018 	& 0.0630 \\
		& PSM1.2		& 68.6286	& 0.0578 		& -0.0127 	& 0.0633 \\
		& PSM1.3		& 77.6452	& 0.0563 		& -0.0109 	& 0.0554 \\
		& PSM2.1		& 57.9343	& 0.0557 		& 0.0013 	& 0.0756 \\
		& PSM2.2		& 67.7502	& 0.0564 		& 0.0015 	& 0.0641 \\
		& PSM2.3		& 67.7401	& 0.0572 		& -0.0121 	& 0.0640 \\
		& PSM3		& 57.0990	& 0.0570 		& 0.0019 	& 0.0767 \\ \hline
2.4		& PSM1.1		& 68.5921	& 0.0540 		& 0.0036 	& 0.0536 \\
		& PSM1.2		& 68.5909	& 0.0585 		& -0.0298 	& 0.0538 \\
		& PSM1.3		& 77.6270	& 0.0555 		& -0.0253 	& 0.0469 \\
		& PSM2.1		& 57.8562	& 0.0551 		& 0.0031 	& 0.0642 \\
		& PSM2.2		& 67.7485	& 0.0526 		& 0.0042 	& 0.0536 \\
		& PSM2.3		& 67.7273	& 0.0579 		& -0.0294	& 0.0546 \\
		& PSM3		& 57.0425	& 0.0548 		& 0.0038 	& 0.0649 \\ \hline
\end{tabular}
\end{table}

\newpage
\noindent \textbf{\large 4. Discussion}

In this study, we systematized a causal inference framework based on a counterfactual model for variance and correlation coefficients.
We demonstrated that the population causal effect, as formulated in this paper, can be identified and estimated conditionally on PS.
These results suggest the applicability of common causal inference methods such as matching, stratification, and inverse probability weighting to our proposed framework.

The findings from the simulation study suggest that PSM can reduce the impact of confounding bias.
As shown in Section 3.1, unadjusted analyses tended to inflate type I errors and produce biased estimates, particularly under heteroscedasticity and strong confounding. 
While including true confounders in the PS model improved accuracy, as seen from PSM1 to PSM3, the results of Section~3.2 suggested that adding irrelevant variables did not necessarily enhance adjustment.
It is important to note that strong confounding may still lead to incorrect conclusions.

This study has several limitations. 
First, the DAG used in the simulations was relatively simple, and more comprehensive simulations with complex structures are warranted. 
Second, alternative matching methods should also be evaluated through simulation. 
Third, other approaches using PSs, such as inverse probability of treatment weighting and stratification, require theoretical clarification and simulation-based assessment. 
These issues are currently being explored as future work.

An important implication of these findings is that PS methods—originally developed for mean comparisons—can be successfully extended to higher-order moments such as variance and correlation. 
This extension is particularly relevant for the DNB, which relies on both components. 
Proper adjustment for confounding is therefore essential to ensure valid group comparisons of DNB indices in clinical studies, where randomization is rarely feasible.
These findings are expected to be applicable in clinical biomarker studies.

\end{document}